\documentclass[twocolumn, amssymb,  amsmath, amsfonts, nobibnotes, a4paper, aps, prd, superscriptaddress, nofootinbib]{revtex4-2}

\usepackage[T3,T1]{fontenc}
\DeclareMathAlphabet{\mathpzc}{OT1}{pzc}{m}{it} 
\usepackage{mathrsfs} 
\usepackage{bm} 
\makeatletter
\@ifclassloaded{beamer}
  { 
    \typeout{UsePackages: Detected beamer}
    \usepackage{tgheros}
    
  }
  { 
    \typeout{UsePackages: Did not detect beamer}
    \ifx\asybeamer\undefined 
    \typeout{UsePackages: Detected article}
    \usepackage[varg]{txfonts} 
    \usepackage{tgtermes} 
    \else 
    \typeout{Fonts: Detected asy for beamer}
    \usepackage{tgheros}
    
    \fi
  }

\usepackage{siunitx}
\usepackage{relsize}
\usepackage{booktabs}
\usepackage{tabularx}
\usepackage[table]{xcolor}

\usepackage{multirow, newfloat}
\usepackage{graphicx} %
\usepackage[x11names,svgnames,rgb]{xcolor} %
\usepackage{xspace}

\usepackage{mathtools}
\DeclareSymbolFontAlphabet{\mathrm}{operators}

\makeatletter %
\newcommand{\ShowFont}{%
  \typeout{The main font is \f@encoding \space \f@family \space %
    \f@series \space \f@shape \space at \f@size pt.}%
  \typeout{The math font sizes are \tf@size pt (main), \sf@size pt %
    (script), and \ssf@size pt (scriptscript).}%
  \typeout{The linewidth is \the\linewidth}} %
\makeatother %

\usepackage[colorlinks]{hyperref}

\usepackage{orcidlink}

\newcommand{\invsec}{\si{\per\second}}

\newcommand{\modelstyle}[1]{\texttt{#1}}

\newcommand{\taylorFtwo}{\mbox{\modelstyle{TaylorF}2}\xspace}
\newcommand{\taylorTtwo}{\mbox{\modelstyle{TaylorT}2}\xspace}

\newcommand{\phXP}{\mbox{\modelstyle{PhenomXP}}\xspace}

\newcommand{\phXPHM}{\mbox{\modelstyle{PhenomXPHM}}\xspace}

\newcommand{\pyEFPE}{\mbox{\modelstyle{pyEFPE}}\xspace}

\newcommand{\losa}{\mbox{\textsc{LoSA}}\xspace}
\newcommand{\los}{\mbox{\textsc{LoS}}\xspace}

\newcommand{\bilby}{\mbox{\textsc{Bilby}}\xspace}
\newcommand{\pbilby}{\mbox{\textsc{Parallel Bilby}}\xspace}

\newcommand{\lalsuite}{\mbox{\textsc{LALSuite}}\xspace}

\newcommand{\lalsimulation}{\mbox{\textsc{LALSimulation}}\xspace}
\newcommand{\dynesty}{\mbox{\textsc{Dynesty}}\xspace}
\newcommand{\numpy}{\mbox{\textsc{NumPy}}\xspace}
\newcommand{\scipy}{\mbox{\textsc{SciPy}}\xspace}
\newcommand{\matplotlib}{\mbox{\textsc{Matplotlib}}\xspace}

\DeclareMathAlphabet{\mathcalstd}{OMS}{cmsy}{m}{n}
\DeclareMathAlphabet{\mathpzc}{OT1}{pzc}{m}{it}

\newcommand{\UCLouvain}{University of Louvain, Centre for Cosmology, Particle Physics and Phenomenology---CP3, \\ Chemin du Cyclotron 2, 1348 Louvain-la-Neuve, Belgium}
\newcommand{\ROB}{Royal Observatory of Belgium, Avenue Circulaire, 3, 1180 Uccle, Belgium}

\definecolor{dodgerblue}{HTML}{1E90FF}
\hypersetup{citecolor=dodgerblue,  urlcolor=dodgerblue}

\definecolor{RED}{HTML}{F5054F}
\definecolor{LIGHT_ORANGE}{HTML}{FDAA48}
\definecolor{DARK_ORANGE}{HTML}{FF5B00}
\definecolor{LIGHT_BLUE}{HTML}{448EE4}
\definecolor{DARK_BLUE}{HTML}{0343df}
\definecolor{LIGHT_GREEN}{HTML}{40C53C}
\definecolor{DARK_GREEN}{HTML}{02590F}

\newcommand{\Hz}{\ensuremath{\mspace{2mu}\text{Hz}}\xspace}

\newcommand{\mediumparallel}{\mathrel{\scalebox{0.85}[0.6]{$\parallel$}}}

\newcommand{\aLOS}{a_{\mediumparallel}}

\newcommand\abs[1]{\ensuremath{\lvert#1\rvert}}

\def \msun  {\rm{M}_\odot}

\begin{document}

\title{Line-of-sight acceleration in compact binaries with higher harmonics and eccentricity}

\author{Soumen Roy\orcidlink{0000-0003-2147-5411}}
\email{soumen.roy@uclouvain.be}
\affiliation{\UCLouvain}
\affiliation{\ROB}

\author{Justin Janquart\orcidlink{0000-0003-2888-7152}}
\affiliation{\UCLouvain}
\affiliation{\ROB}

\begin{abstract}

Direct detections of gravitational waves provide a unique opportunity to probe the astrophysical origin of compact binary mergers. The formation channels of these systems remain highly debated, and a fraction may originate in dynamical environments or active galactic nuclei. Binaries formed in such environments are expected to experience line-of-sight acceleration from their surroundings, which can imprint characteristic signatures on the observed gravitational-wave signal.
Here, we re-derive the line-of-sight acceleration effects and implement them in state-of-the-art quasi-circular waveform models with precession and higher-order modes. We also implement the corrections in eccentric waveform models, applying them consistently to all contributing harmonics.
Using this model, we investigate the impact of these effects on the inference of line-of-sight acceleration and analyze a few interesting GWTC events observed during the third observing run of LIGO and Virgo. We find no substantial evidence for line-of-sight acceleration in these events.  We also show that an inconsistent treatment of line-of-sight acceleration between higher harmonics can lead to biased conclusions. Our model provides a robust framework for uncovering line-of-sight acceleration in current and future gravitational-wave observations, enabling more accurate probes of environmental signatures in compact-binary formation.

\end{abstract}
\maketitle

\section{Introduction}

The LIGO–Virgo–KAGRA (LVK) gravitational-wave (GW) detectors network~\cite{LIGOScientific:2014pky, VIRGO:2014yos, KAGRA:2020tym} has reported over 390 events through the second part of the fourth observing run~\cite{LIGOScientific:2026wfs, LIGOScientific:2025slb, KAGRA:2021vkt, LIGOScientific:2020ibl, LIGOScientific:2018mvr, Nitz:2021zwj, Wadekar:2023gea}, all of which are produced by mergers of compact binaries. The fifth observing run is expected to deliver a significant boost in sensitivity~\cite{KAGRA:2013rdx}. These detections, comprising mergers of binary black hole (BBH), binary neutron star (BNS), and neutron star–black hole (NSBH) systems, have opened new avenues for studying compact objects and the environments in which they form and evolve.

The growing catalog of compact-binary mergers has enabled population-level studies of masses, spins, and merger rates, providing new constraints on compact-object formation~\cite{LIGOScientific:2025pvj, KAGRA:2021duu}. However, disentangling formation channels remains challenging because multiple scenarios can produce overlapping distributions in mass and spin~\cite{KAGRA:2021duu, Antonelli:2023gpu}. In particular, the relative contributions of isolated binary evolution and dynamically driven mergers remain uncertain. The latter can occur in a variety of environments, including dense stellar systems such as nuclear star clusters~\cite{Rodriguez:2016kxx, Mapelli:2021gyv}, hierarchical triples or higher-order multiples~\cite{Antonini:2017ash, Silsbee:2016djf, Liu:2018nrf}, as well as gas-rich environments such as active galactic nucleus (AGN) disks~\cite{Bartos:2016dgn, McKernan:2020lgr, Rowan:2022ehz}. Compact objects embedded in AGN disks may migrate close to the central supermassive black hole (SMBH) and become trapped there~\cite{Peng:2021vzr}.

Orbital eccentricity is a promising discriminator between isolated and dynamical formation scenarios, since dynamically assembled binaries can retain measurable eccentricity when entering the detector band~\cite{Zevin:2021rtf, DallAmico:2023neb}. Nevertheless, eccentricity alone cannot uniquely identify which dynamical channel produced a given merger~\cite{Stegmann:2025shr}. Recent studies have therefore proposed joint mass--spin--eccentricity signatures as a more powerful probe of formation channels~\cite{Fumagalli:2023hde, Stegmann:2025shr}. However, eccentricity can be substantially radiated away before the signal reaches the sensitive bandwidth of ground-based detectors, reducing its discriminating power~\cite{Peters:1964zz, Tucker:2021mvo}. Moreover, degeneracies among mass, spin, and eccentricity in the waveform parameter space can make eccentricity--spin disentanglement difficult~\cite{CalderonBustillo:2020xms, Romero-Shaw:2025vbc, Divyajyoti:2025cwq, Tibrewal:2026jci}. These limitations motivate the development of complementary approaches that search for direct environmental signatures encoded in the GW signal.

Line-of-sight acceleration (\losa) of the binary’s center-of-mass motion induced by an external potential or a nearby perturber, provides a promising probe of dynamical formation~\cite{Takahashi:2004yr, Yunes:2010sm, Meiron:2016ipr, Bonvin:2016qxr, Inayoshi:2017hgw, Tamanini:2019usx, Vijaykumar:2023tjg, Tiwari:2025aec, Tiwari:2024pvb, Lazarow:2024gdn, Gera:2025ugl}. This acceleration can arise in hierarchical triples, dense stellar environments, or AGN-assisted scenarios~\cite{Meiron:2016ipr, Inayoshi:2017hgw, Samsing:2024syt, Hendriks:2024zbu, Tiwari:2023cpa, Zwick:2025wkt, Tagawa:2025tfd}, and imprints a characteristic frequency-dependent distortion in the waveform. For the latter, another particular signature could be that of lensing, due to the emitted GW being lensed by the central black hole~\cite{Leong:2024nnx}. Should such events be detected, they would be excellent targets to search for \losa.

Crucially, unlike a constant redshift (from cosmology) or a constant line-of-sight (\los) velocity, which is degenerate with a simple rescaling of detector-frame masses, \losa{} induces a \emph{time-dependent} Doppler modulation. This time dependence cannot be absorbed into a constant mass rescaling and instead leaves a characteristic imprint on the waveform phasing. Working to linear order in the \los redshift, the leading-order correction to the Fourier-domain phase can be written as~\cite{Takahashi:2004yr, Bonvin:2016qxr}
\begin{equation}
\Delta\Psi(f) = \frac{25}{65536\,\eta^2}\frac{G M}{c^3}
\frac{\aLOS}{c} \, v^{-13},
\label{eq:losa_leading_phase}
\end{equation}
where $\aLOS$ is the acceleration of the source along \los, $M$ and $\eta$ denote the total mass and symmetric mass ratio of the binary, and $v = (\pi G M f/c^3)^{1/3}$ is the post-Newtonian (PN) velocity parameter (often just called the PN expansion parameter). This leading order correction enters at $-4$PN order relative to the leading vacuum phase and is enhanced for low-mass systems that spend longer in band.

The \losa effect on GW signals has been explored in several contexts. Focusing on cosmic acceleration, Takahashi and Nakamura~\cite{Takahashi:2004yr} provided an early frequency-domain treatment of waveform distortions induced by a time-dependent redshift, which is conceptually similar to \losa effects from center-of-mass motion. Bonvin et al.~\cite{Bonvin:2016qxr} formulated \losa as a time-dependent Doppler modulation and derived the leading-order correction to the Fourier-domain phase, emphasizing its impact on parameter inference, particularly for long-lived inspirals. Tamanini et al.~\cite{Tamanini:2019usx} extended the treatment to higher-order corrections to this leading contribution, up to 1.5PN order and presented forecasts for constraints on $\aLOS/c$ from stellar-mass binaries observed with LISA, where the long observation times could significantly enhance the relevance of \losa. Vijaykumar et al.~\cite{Vijaykumar:2023tjg} derived the correction to the quasi-circular inspiral phase through $3$PN order (to linear order in $\aLOS/c$), and also provided the leading-order amplitude correction. A follow-up study by Tiwari et al.~\cite{Tiwari:2025aec} implemented \losa waveform corrections for parameter estimation, computed the phase correction up to $3.5$PN order (to linear order in $\aLOS/c$), and carried out a systematic study of potential biases and modeling uncertainties relevant for \losa inference. Another recent study by Lazarow et al.~\cite{Lazarow:2024gdn} constructed an analytic frequency-domain model by treating the (approximately constant) center-of-mass acceleration as a perturbation to the \taylorFtwo approximant, and computed the waveform phase through $3$PN order at linear order in $\aLOS/c$), providing Fisher forecasts for current and next-generation detectors. A complementary approach by Chamberlain et al.~\cite{Chamberlain:2018snj} derived Doppler shifts using a stationary-phase treatment in the frequency domain, incorporating both phase and amplitude corrections.

Most recently, \losa{} has been included as part of the GWTC-4.0 tests of General Relativity (GR)~\cite{LIGOScientific:2026fcf}, using a quasi-circular implementation based on the dominant quadrupole treatment and following the waveform-correction prescription developed in~\cite{Vijaykumar:2023tjg, Tiwari:2025aec}. In that analysis, the \losa study was restricted to inspiral-dominated systems, including selection cuts on the total mass and only considering near-equal-mass events $q =m_2/m_1 \ge 0.25$ to avoid biases due to higher-order multipoles when using a dominant-quadrupole \losa implementation; as a result, only a subset of events satisfies the selection criteria. The GWTC-4.0 analysis reports no statistically significant evidence for \losa, and provides 90\% credible intervals on $\aLOS/c$ for that subset of events.

The existing \losa treatments have primarily focused on quasi-circular binaries and derive the waveform correction by propagating the Doppler-induced redshift through the PN evolution and reconstructing the Fourier-domain phase. While this is fully adequate for the dominant mode, extending it to higher harmonics or eccentric waveforms is cumbersome because the mode-dependent frequency mapping must be treated separately for each harmonic. 

In this work, we extend the treatment of \losa beyond the dominant quasi-circular quadrupole by incorporating higher harmonics and eccentric compact binaries. Our formulation is based on the integrated time delay induced by the \los redshift and expresses the Fourier-domain phase modification in a closed form in terms of the stationary-phase time. In this form, the \losa effect can be applied mode by mode by replacing the quadrupolar stationary-phase time with the corresponding harmonic-dependent quantity $t_{\ell m}(f)$, and can likewise be generalized to eccentric waveforms by using the appropriate stationary times for each contributing harmonic. This provides a unified framework in which \losa corrections can be implemented consistently across waveform models with richer harmonic structure.

The remainder of this paper is organized as follows. In Sec.~\ref{sec:method}, we summarize the \losa waveform correction and describe its implementation in higher-harmonic and eccentric waveform models. In Sec.~\ref{sec:injections}, we present injection studies quantifying measurability and biases when higher harmonics or eccentricity are omitted. In Sec.~\ref{sec:real_events}, we analyze selected LVK events, and we conclude in Sec.~\ref{sec:conclusion} with a discussion of implications and prospects for future detectors.

\section{Phase and Amplitude Corrections from Line-of-Sight Acceleration}
\label{sec:method}

We derive the leading-order correction to the frequency-domain GW phase induced by a time-dependent Doppler shift arising from the \los acceleration of the binary’s center of mass. Throughout this work, the \los redshift is treated to linear order, assuming $|z_\ell|\ll 1$. The \los redshift is defined as $z_\ell(t_{\rm src}) \equiv v_{\mediumparallel}(t_{\rm src})/c$, where $v_{\mediumparallel} (t_{\rm src})$ is the \los velocity of the binary’s center of mass. For constant acceleration $\aLOS$, the redshift evolves as
\begin{equation}
z_\ell(t_{\rm src}) = z_0 + \Gamma t_{\rm src},
\qquad
\Gamma \equiv \frac{\aLOS}{c},
\end{equation}
where $z_0$ represents a constant velocity contribution.

\subsection{Phase Correction}

We first consider the waveform in the source frame. The dominant mode of time-domain waveform can be written as $h_{\rm src}(t_{\rm src}) \simeq A(t_{\rm src}) e^{i\phi(t_{\rm src})}$. In the stationary phase approximation (SPA), the Fourier transform of the waveform is dominated by stationary times satisfying
\begin{equation}
\frac{d\phi}{dt_{\rm src}}(t_{\rm src}) = 2\pi f_{\rm src},
\label{SPA_src}
\end{equation}
where $f_{\rm src}$ is the GW frequency in the source frame. For compact binaries, the GW phase is related to the orbital phase $\Phi$ through $\phi(t)=2\Phi(t)$. Since $\dot{\phi}(t)=2\Omega(t)$, with $\Omega$ the orbital frequency, Eq.~(\ref{SPA_src}) implies $\Omega(t_{\rm src}) = \pi f_{\rm src}$. Thus $t_{\rm src}$ represents the emission time at which a GW of source-frame frequency $f_{\rm src}$ is generated. In PN theory, it is convenient to introduce the velocity parameter $v=(\pi M f_{\rm src})^{1/3}$. In the \taylorTtwo approximant~\cite{Damour:2000zb}, the time-to-coalescence is expressed as a function $t(v)$. Equivalently, in the SPA used in \taylorFtwo~\cite{Buonanno:2009zt}, the stationary time is given by the derivative of the frequency-domain phase $(\Psi_{\rm src})$ with respect to the source-frame frequency,

\begin{equation}
t(f_{\rm src}) =
\frac{1}{2\pi}\frac{d\Psi_{\rm src}}{df_{\rm src}},
\end{equation}
which coincides with the PN time-to-coalescence in the \taylorTtwo construction. Hence, $t_{\rm src}=t(f_{\rm src})$. We measure source times relative to the coalescence time $t_c$, so that $t_{\rm src}=0$ at merger and $t_{\rm src}<0$ during the inspiral.

We now introduce the Doppler mapping between source and detector frames. The detector time $t_{\rm det}$ and the source time $t_{\rm src}$ are related by
\begin{equation}
\frac{dt_{\rm det}}{dt_{\rm src}}
= 1 + z_\ell(t_{\rm src}).
\end{equation}
Integrating this relation gives the cumulative delay
\begin{equation}
\label{eq:time_delay}
\Delta t(t_{\rm src})
\equiv t_{\rm det}-t_{\rm src}
=
\int^{t_{\rm src}} z_\ell(t')\,dt' =
z_0 t_{\rm src} + \frac{1}{2}\Gamma t_{\rm src}^2 .
\end{equation}
The term proportional to $z_0$ corresponds to a constant Doppler shift and is degenerate with a rescaling of intrinsic parameters such as the masses and the coalescence time. The acceleration term, however, produces a frequency-dependent distortion of the waveform. 

This effect can be interpreted as a time-dependent R{\o}mer delay: the line-of-sight motion of the binary’s center of mass changes the light-travel time between emission and detection, producing an accumulated arrival-time shift that translates into a frequency-dependent phase distortion. Equivalent R{\o}mer-delay formulations have been used extensively in the context of externally perturbed binaries and hierarchical triples~\cite{Samsing:2024syt, Hendriks:2024zbu, Hendriks:2024gpp}. A related frequency-domain treatment of Doppler-induced corrections within the stationary-phase approximation was presented in~\cite{Chamberlain:2018snj}.

Since the Fourier-domain phase is defined relative to the
coalescence time, $\Psi(f) = 2\pi f t_c - \phi_c + \psi_{\rm PN}(f)$, the relevant delay is the cumulative redshift accumulated between the emission time and merger,
\begin{equation}
\Delta t(f)
=
\int_{t_{\rm src}(f)}^{0}
z_\ell(t')\,dt'.
\end{equation}
The corresponding correction to the Fourier-domain phase is
\begin{equation}
\Delta\Psi(f)
=
-2\pi f\,\Delta t(f).
\end{equation}
Retaining only the acceleration term in Eq.~\eqref{eq:time_delay} yields
\begin{equation}
\label{eq:losa_phasecorr}
\Delta\Psi(f) = -\pi \: \Gamma f\, t(f)^2 
\end{equation}
We note that, since the Fourier-domain phase is referenced to coalescence, the acceleration-induced phase correction vanishes at the merger frequency. The constant velocity contribution proportional to $z_0$ is absorbed into intrinsic parameters, while only the acceleration term produces a frequency-dependent phase distortion that is largest at low frequencies and decreases toward coalescence.

The above expression can be straightforwardly generalized to higher-order modes. For a given $(\ell,m)$ mode, the stationary time entering the phase correction is replaced by the corresponding mode-dependent SPA time $t_{\ell m}(f)$. The frequency domain phase for an arbitrary mode is related to the dominant quadrupole mode by $\psi_{\ell m}(f) = m/2 \,\psi_{22}\!\left(2f/m\right)$, which implies $t_{\ell m}(f) = t_{22}\!\left(2f/m\right)$. The correction to phase for the $(\ell,m)$ mode therefore becomes,
\begin{equation}
\label{eq:losa_phasecorr_hom}
\Delta\Psi_{\ell m}(f) = -\pi\,\Gamma\, f\, t_{\ell m}(f)^2.
\end{equation}
Higher-order modes therefore probe different portions of the inspiral through their mode-dependent stationary times $t_{\ell m}(f)$, typically corresponding to earlier inspiral times than the dominant $(2,2)$ mode at a fixed observed frequency. As a result, the associated \losa{} phase corrections can differ appreciably across modes when the higher-order-mode content is non-negligible. 

\subsection{Amplitude Correction}

In addition to the phase, the \losa acceleration also modifies the Fourier-domain amplitude through the mapping between source-frame and detector-frame variables. We derive this correction following the same setup used for the phase. The time-domain spherical-harmonic modes are written in terms of source-frame quantities as
\begin{equation}
    h_{\ell m}(t_{\rm src}) = A_{\ell m}(x_{\rm src})\, e^{i\phi_{\ell m}(t_{\rm src})},
\end{equation}
where \(x_{\rm src} \equiv \left(2\pi M f_{\rm src}/m\right)^{2/3}\)
is the source-frame PN expansion parameter for a generic \(m\)-mode. We similarly define the detector-frame PN parameter by \(x \equiv \left(2\pi M f/m\right)^{2/3}\),
where \(f\) is the observed Fourier frequency. In the absence of a time-dependent redshift, these two quantities coincide.

In the stationary-phase approximation, the Fourier-domain amplitude can be written as
\begin{equation}
\label{eq:SPA_amp}
A_{\ell m}^{\rm SPA}(f) = A_{\ell m}(x_{\rm src})\, \sqrt{\frac{2\pi}
{m\,(3/2)\sqrt{x}\,\dot{x}}},
\end{equation}
where \(\dot{x}\equiv dx/dt_{\rm det}\)
is the detector-frame chirp rate, evaluated at the mode-dependent stationary time \(t_{\rm src}=t_{\ell m}(f)\).
For a given observed frequency \(f\), different harmonics correspond to different orbital frequencies and therefore to different stationary times through \(x_m(f)=\left(2\pi M f/m\right)^{2/3}\).
Following the phase derivation, the \losa induces a time-dependent redshift that relates source-frame and detector-frame frequencies according to
\begin{equation}
\label{eq:xrdshift}
 f_{\rm src} = f\,[1+z_\ell(t_{\rm src})],
\qquad
x_{\rm src} = x\,[1+z_\ell(t_{\rm src})]^{2/3}.
\end{equation}
Differentiating Eq.~\eqref{eq:xrdshift} with respect to \(t_{\rm det}\) and using Eq.~\eqref{eq:time_delay} yields
\begin{equation}
\label{eq:dxdt}
\frac{dx}{dt_{\rm det}}
=
(1+\Gamma t_{\rm src})^{-5/3}
\frac{dx_{\rm src}}{dt_{\rm src}}
-
\frac{2}{3}
\frac{\Gamma x}
{(1+\Gamma t_{\rm src})^{2}},
\end{equation}
where \(d/dt_{\rm det}=[1/(1+\Gamma t_{\rm src})]\,d/dt_{\rm src}\). At Newtonian order, the intrinsic source-frame evolution is
\begin{equation}
\label{eq:dxdtN}
\left(
\frac{dx_{\rm src}}{dt_{\rm src}}
\right)_{\!N}
=
\frac{64\eta}{5M}x_{\rm src}^{5}.
\tag{15}
\end{equation}
Substituting Eq.~\eqref{eq:dxdtN} into Eq.~\eqref{eq:dxdt} and expanding to first order in \(\Gamma\) gives
\begin{equation*}
\frac{dx}{dt_{\rm det}} = K x^5 \left[
1+\frac{5}{3}\Gamma t_{\rm src}-\frac{2\Gamma}{3Kx^4} \right],
\end{equation*}
where \(K\equiv 64\eta/(5M)\). The Newtonian stationary time for the \((\ell,m)\) mode is
\[
t_{\ell m}(f) = -\frac{1}{4Kx_m^4}
= -\frac{5M}{256\eta}x_m^{-4},
\]
where $t_{\ell m}(f)<0$ during the inspiral. Therefore,
\[ -\frac{2\Gamma}{3Kx_m^4} = \frac{8}{3}\Gamma t_{\ell m}(f). \]
Combining this term with the first contribution proportional to
$\Gamma$, we obtain
\begin{equation}
\frac{dx}{dt_{\rm det}} = \left(\frac{dx}{dt}\right)_{\!N}
\left[
1+\frac{13}{3}\Gamma t_{\ell m}(f)
\right].
\tag{16}
\end{equation}
Here the coefficient \(13/3\) is common to all harmonics, while the stationary time \(t_{\ell m}(f)\), or equivalently \(x_m(f)\), is mode dependent.

At fixed observed frequency $f$, the detector-frame quantity $x=x_m(f)$ is fixed. However, Eq.~\eqref{eq:SPA_amp} shows that the Fourier-domain amplitude receives two distinct \losa corrections. The first arises from the intrinsic time-domain amplitude, which is evaluated at the source-frame quantity \(x_{\rm src}\), and the second comes from the modification of the detector-frame chirp rate appearing in the SPA factor \(\dot{x}^{-1/2}\). Thus,
\begin{equation}
\frac{\delta A_{\ell m}^{\rm SPA}}
{A_{\ell m}^{\rm SPA}}
=
\frac{\delta A_{\ell m}^{\rm TD}}
{A_{\ell m}^{\rm TD}}
-
\frac{1}{2}
\frac{\delta\dot{x}}{\dot{x}}.
\end{equation}

To evaluate the first contribution, we define the logarithmic slope of the time-domain mode amplitude,
\begin{equation}
\alpha_{\ell m}(x)
\equiv
\frac{d\ln A_{\ell m}}{d\ln x}.
\end{equation}
From Eq.~\eqref{eq:xrdshift},
\begin{equation}
x_{\rm src} = x
\left[
1+\frac{2}{3}\Gamma\, t_{\ell m}(f)
\right]
+\mathcal{O}(\Gamma^2).
\end{equation}
Expanding \(A_{\ell m}(x_{\rm src})\) about \(x\), the source-amplitude contribution is
\begin{equation}
\frac{\delta A_{\ell m}^{\rm TD}}
{A_{\ell m}^{\rm TD}} =
\frac{2}{3}\alpha_{\ell m}(x)\,\Gamma\,t_{\ell m}(f).
\end{equation}
The second contribution follows directly from Eq.~(16). Since the SPA amplitude is proportional to \(\dot{x}^{-1/2}\),
\begin{equation}
-\frac{1}{2}\frac{\delta\dot{x}}{\dot{x}}
=
-\frac{13}{6}\Gamma\,t_{\ell m}(f).
\end{equation}

The total leading-order fractional amplitude correction is therefore
\begin{equation}
\label{eq:losa_amplitudecorr}
\frac{\delta A_{\ell m}^{\rm SPA}} {A_{\ell m}^{\rm SPA}} =
\left[ \frac{2}{3}\alpha_{\ell m}(x) - \frac{13}{6} \right]
\Gamma\,t_{\ell m}(f).
\end{equation}
This expression includes both the Doppler shift of the intrinsic time-domain waveform amplitude and the modification of the SPA curvature. The former depends on the leading frequency scaling of each individual mode through \(\alpha_{\ell m}\), while the latter is determined by the acceleration-induced correction to the detector-frame chirp rate. For the leading Newtonian time-domain mode amplitudes, the generic scaling is given by~\citep{Damour:2008gu},
\begin{equation}
A_{\ell m}^{\rm N}(x)\propto x^{(\ell+\epsilon_{\ell m})/2},
\end{equation}
where $\epsilon_{\ell m}=0$ for mass-type multipoles where $\ell+m$ is even and $\epsilon_{\ell m}=1$ for current-type multipoles where $\ell+m$ is odd. Therefore, $\alpha_{\ell m}^{\rm N} = (\ell+\epsilon_{\ell m})/2$ and for the $(2,2)$ mode, the leading order amplitude correction is,
\begin{equation}
\frac{\delta A_{22}^{\rm SPA}}{A_{22,\,{\rm lead}}^{\rm SPA}}
=
\frac{60}{2048}\frac{\Gamma M}{\eta}x^{-4}.
\end{equation}

\subsection{Implementation in \phXPHM model}
The \phXPHM waveform model is a frequency-domain inspiral--merger--ringdown model for generically precessing BBHs~\cite{Pratten:2020ceb, Garcia-Quiros:2020qpx}. It incorporates higher-order harmonics by constructing the waveform in a co-precessing frame and then rotating it to the inertial observer frame~\cite{Hannam:2013oca}. In this way, the dominant precession effects are captured through a time- (or frequency-) dependent rotation, while the underlying mode structure is modeled in a frame approximately aligned with the instantaneous orbital angular momentum.

In this model, the GW polarizations in the inertial $J$-frame, $\tilde{h}_{+,\times}^{J}(f)$, are written in terms of the spherical harmonic modes $\tilde{h}_{\ell m}^{L}(f)$ defined in the co-precessing $L$-frame as
\begin{align}
\label{eq:polarizations}
    &\tilde{h}_{+,\times}^{J} (f>0) = \frac{1}{2} \sum_{\ell \geq 2} \sum_{m^{\prime} > 0}^{\ell} \tilde{h}^{L}_{\ell  -m^{\prime}} (f) e^{ i m' \gamma}
    \nonumber\\
    &\quad \times \sum_{m=-\ell}^{\ell}\left[ e^{-i m \alpha} d^{\ell}_{m -m'}(\beta)\, _{-2}Y_{\ell m}
    \:\pm\:
    (-1)^{\ell} e^{-i m \alpha} d^{\ell}_{m m'}(\beta)\, _{-2}Y_{\ell m}^{\ast} \right],
\end{align}
where the $\pm$ signs on the right-hand side correspond to the $+$ and $\times$ polarizations, respectively. Effectively, the above equation can be written as 
\begin{equation}
    \label{eq:polarizations_H}
    \tilde{h}_{+,\times}^{J}(f)=\sum_{\ell,m} H^{+,\times}_{\ell m}(f) \, .
\end{equation}
Here $d^{\ell}_{mm'}(\beta)$ are the real Wigner \(d\)-matrices, $(\alpha,\beta,\gamma)$ are the Euler angles that rotate the co-precessing $L$-frame into the inertial $J$-frame, and $_{-2}Y_{\ell m}$ are the spin-weighted spherical harmonics of spin weight $-2$.

For the \losa implementation, we first extract the effective mode contributions $H^{+,\times}_{\ell m}$ from the \lalsimulation subpackage in \lalsuite software~\cite{lalsuite}. We then apply the \losa correction to each mode individually using Eqs.~\eqref{eq:losa_phasecorr_hom} and \eqref{eq:losa_amplitudecorr}, and finally reconstruct the corrected polarizations in the inertial frame,
\begin{equation}
\tilde{h}_{+,\times}^{J}(f)=\sum_{\ell,m} H^{+,\times}_{\ell m}(f) \, \left(1 + \frac{\delta A_{\ell m}^{\rm SPA}}{A_{\ell m}^{\rm SPA}}\right) \, e^{- i \Delta\Psi_{\ell m}(f)} \, .
\end{equation}
To compute $t_{\ell m}(f)$, we employ the PN phasing coefficients of \taylorFtwo model up to 3.5PN order, as implemented in \lalsuite software~\cite{lalsuite}.

\subsection{Implementation in \pyEFPE model}
The \pyEFPE waveform model is a frequency-domain inspiral model for precessing-eccentric compact binaries~\cite{Morras:2025nlp}, constructed within the Efficient Fully Precessing Eccentric (EFPE) waveform paradigm~\cite{Klein:2021jtd, Arredondo:2024nsl, Morras:2025nlp}. In this framework, the orbital motion is described using the post-Newtonian formalism together with the quasi-Keplerian parameterization~\cite{Damour:1985, Damour:1986}. The spin precession is incorporated through a multiple-scale analysis~\cite{Klein:2013qda, Gerosa:2023xsx}, assuming that the spins and orbital angular momentum evolve slowly. The waveform is assembled from stationary-phase quantities for each contributing harmonic together with a shifted uniform asymptotics (SUA) treatment of the amplitude~\cite{Klein:2014bua}. At present, \pyEFPE is an inspiral-only model and does not include higher-order modes or the merger-ringdown part of the signal.

In this waveform model, the stationary-phase quantities $t_i^{\rm SPA}$, $T_i^{\rm SPA}$, and $\psi_i^{\rm SPA}$ are computed for each contributing harmonics before waveform assembly.
    \begin{widetext}
\begin{equation}
\label{eq:pyEFPEwf}
\tilde{h}_{+,\times}(f) = \underbrace{
\sum_i \sqrt{2\pi}\: T_i^{\text{SPA}} \,
e^{i \left( 2\pi f t_i^{\text{SPA}} - \phi(t_i^{\text{SPA}}) - \pi/4 \right)}}_{\text{\normalsize Fourier modes}} \quad
\underbrace{\sum_{k = -k_{\max}}^{k_{\max}} a_{k, k_{\max}} \,
\mathcal{A}^{+,\times}_{2, m_i, n_i}\left(t_i^{\text{SPA}} + k T_i^{\text{SPA}} \right)}_{\text{\normalsize SUA correcttion to amplitude }  \mathlarger{(A_i^{\text{SUA}})} }
\end{equation}
\begin{equation}
\label{eq:pyEFPE}
\text{Effectively,}\quad \tilde{h}(f) = \sum_i \tilde{h}_i(f) = \sum_i A_i^{\text{SUA}} \sqrt{2\pi}\: T_i^{\text{SPA}} \exp\left\{i \, \psi^{\text{SPA}}_i \right\}
\end{equation}
\begin{equation}
    \text{where, }\quad \left.\frac{d\phi(t)}{dt}\right|_{t = t^{\text{SPA}}} = 2\pi f, \qquad T^{\text{SPA}} = \left| \frac{d^2 \phi(t)}{dt^2} \right|^{-1/2}_{t = t^{\text{SPA}}}, \, \text{and} \qquad \psi^{\text{SPA}}_i =  2\pi f t_i^{\text{SPA}} - \phi(t_i^{\text{SPA}}) - \pi/4 \, .
\end{equation}
\end{widetext}
Explicitly, the \pyEFPE waveform generator first computes \(t_i^{\text{SPA}}\), \(T_i^{\text{SPA}}\), \(\psi_i^{\text{SPA}}\), and \(A_i^{\text{SUA}}\), and then substitutes them into Eq.~\eqref{eq:pyEFPE}.

\begin{figure*}
  \centering%
\includegraphics[width=0.95\linewidth]{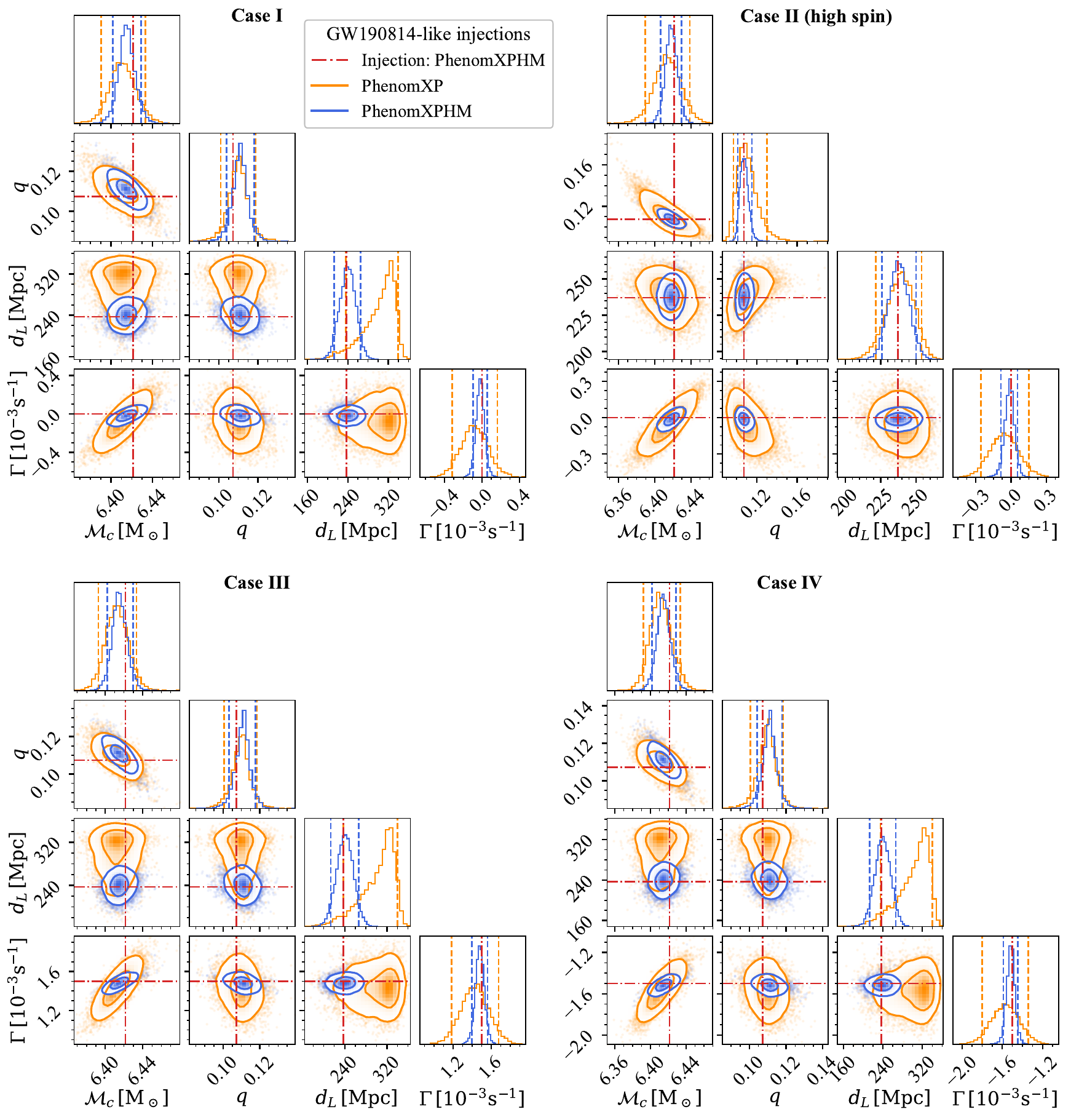}%
  \caption{Systematic biases in \losa inference due to missing higher harmonics in the recovery waveform, shown for GW190814-like injections. The corner plot compares the recovered posteriors with \phXPHM (blue) and \phXP (orange) models, where the injections are generated using the \phXPHM model. The red dashed lines indicate the injected values. The vertical dashed-dot lines in 1D histograms represent the 90\% credible interval. The two consecutive contours in the 2D joint plot correspond to 50\% and 90\% credible regions. When \losa is present and related corrections are neglected in higher-order modes of the recovery waveform, biases are observed.}
  \label{fig:gw190814_inj12}
\end{figure*}

The effect of \losa is incorporated by modifying these stationary-phase quantities according to Eq.~\eqref{eq:losa_phasecorr}. For constant acceleration, the phase correction implemented in the model is, 
\begin{equation}
\label{eq:losa_ecc}
\Delta\psi_i^{\rm SPA}(f) = -\pi\, \Gamma\, f\, \left(t_i^{\rm SPA}\right)^2
\end{equation}
with the corresponding amplitude correction applied multiplicatively through the \losa-corrected \(T_i^{\rm SPA}\).

\begin{table*}[ht]
    \centering
    \setlength{\tabcolsep}{4pt}
    \rowcolors{1}{}{gray!15}
    \begin{tabularx}{0.94\textwidth}{r c c c c c c}
        \toprule[1pt]
        \toprule[1pt]
         & $\mathcal{M}_c\:[\msun]$ & $q$ & $\chi_{\rm eff}$ & $\chi_{\rm p}$ & $d_L$ [Mpc] & $\Gamma\:[10^{-3} \invsec]$ \\
        \midrule[0.5pt]
        \multirow{2}{*}{\cellcolor{gray!0}\rotatebox[origin=c]{90}{Case I}}
          &  6.42  & 0.11 & 0.03 & 0.06 & 237.14 & 0.0 
         \vspace{1mm}\\
         & ${6.41}^{+0.01}_{-0.01}  \,\Big\backslash {6.41}^{+0.02}_{-0.02}$ & ${0.11}^{+0.01}_{-0.01}  \,\Big\backslash {0.11}^{+0.01}_{-0.01}$ & ${0.0}^{+0.06}_{-0.05}  \,\Big\backslash {0.01}^{+0.07}_{-0.06}$ & ${0.03}^{+0.03}_{-0.03}  \,\Big\backslash {0.04}^{+0.04}_{-0.03}$ & ${240.0}^{+25.1}_{-27.1}  \,\Big\backslash {309.9}^{+34.8}_{-58.8}$ & ${-0.02}^{+0.07}_{-0.08}  \,\Big\backslash {-0.08}^{+0.23}_{-0.25}$ \vspace{3mm}\\
         \multirow{2}{*}{\cellcolor{gray!0}\rotatebox[origin=c]{90}{Case II}}
        &  ${6.42} $ & ${0.11} $ & ${0.46} $ & ${0.53} $ & ${237.14} $ & ${0.0} $ \vspace{1mm} \\

& ${6.42}^{+0.01}_{-0.01}  \,\Big\backslash {6.41}^{+0.02}_{-0.02}$ & ${0.11}^{+0.01}_{-0.01}  \,\Big\backslash {0.11}^{+0.02}_{-0.02}$ & ${0.45}^{+0.03}_{-0.03}  \,\Big\backslash {0.44}^{+0.07}_{-0.07}$ & ${0.53}^{+0.05}_{-0.05}  \,\Big\backslash {0.55}^{+0.11}_{-0.1}$ & ${237.7}^{+11.8}_{-12.2}  \,\Big\backslash {239.1}^{+15.5}_{-16.7}$ & ${-0.01}^{+0.07}_{-0.07}  \,\Big\backslash {-0.06}^{+0.21}_{-0.2}$ \vspace{3mm} \\
         \multirow{2}{*}{\cellcolor{gray!0}\rotatebox[origin=c]{90}{Case III}}
        &  ${6.42} $ & ${0.11} $ & ${0.03} $ & ${0.06} $ & ${237.14} $ & ${1.5} $ \vspace{1mm} \\
& ${6.42}^{+0.01}_{-0.01}  \,\Big\backslash {6.41}^{+0.02}_{-0.02}$ & ${0.11}^{+0.01}_{-0.01}  \,\Big\backslash {0.11}^{+0.01}_{-0.01}$ & ${0.0}^{+0.06}_{-0.05}  \,\Big\backslash {0.01}^{+0.07}_{-0.06}$ & ${0.03}^{+0.04}_{-0.02}  \,\Big\backslash {0.04}^{+0.03}_{-0.03}$ & ${240.5}^{+26.7}_{-26.6}  \,\Big\backslash {311.3}^{+34.4}_{-60.0}$ & ${1.48}^{+0.08}_{-0.08}  \,\Big\backslash {1.43}^{+0.26}_{-0.24}$ \vspace{3mm} \\
         \multirow{2}{*}{\cellcolor{gray!0}\rotatebox[origin=c]{90}{Case IV}}
        &  ${6.42} $ & ${0.11} $ & ${0.03} $ & ${0.06} $ & ${237.14} $ & ${-1.5} $  \vspace{1mm} \\
& ${6.41}^{+0.01}_{-0.01}  \,\Big\backslash {6.41}^{+0.02}_{-0.02}$ & ${0.11}^{+0.01}_{-0.01}  \,\Big\backslash {0.11}^{+0.01}_{-0.01}$ & ${0.0}^{+0.05}_{-0.05}  \,\Big\backslash {0.01}^{+0.07}_{-0.06}$ & ${0.03}^{+0.03}_{-0.03}  \,\Big\backslash {0.04}^{+0.04}_{-0.03}$ & ${239.7}^{+27.5}_{-25.6}  \,\Big\backslash {310.2}^{+36.5}_{-60.9}$ & ${-1.52}^{+0.07}_{-0.07}  \,\Big\backslash {-1.57}^{+0.22}_{-0.24}$ \vspace{2mm}\\
        \bottomrule[1pt]
        \bottomrule[1pt]
    \end{tabularx}
    \caption{Summary of the \losa inference for GW190814-like injection analyses are tabulated for various parameters. For each injection case, the first row (shaded) contains injection values and the second row shows the 90\% credible intervals obtained from the analyses with \mbox{\phXPHM$\backslash$\phXP} model. The other injected extrinsic parameters are set for all the cases as follows: polarization angle of 2.3, geocentric GPS time of 1264316116.4 seconds, sky location at right ascension 5.6 and declination 0.04, and a coalescence phase of 0.41.
    }
    \label{tab:GW190814res}
\end{table*}


\section{Impact of higher harmonics}
\label{sec:injections}
In this section, we investigate the impact of higher harmonics in the \losa analysis by comparing waveform models that either include or omit them. We consider injections for which higher-order modes contribute significantly to the signal and perform the analyses using the \phXP~\cite{Pratten:2020ceb} and \phXPHM models. The latter includes both precession and higher-order modes, whereas the former includes precession but neglects higher-order modes.

We perform Bayesian inference using the \pbilby library based on the \bilby package~\cite{Smith:2019ucc, Ashton:2018jfp, Romero-Shaw:2020owr}, with the nested sampling algorithm \dynesty~\cite{Speagle:2019ivv, skilling2006}. We use 1500 live points, the \texttt{acceptance-walk} sampling method, and set the acceptance parameter to \texttt{naccept}=60\footnote{Further details on the \dynesty sampler settings are available at \url{https://bilby-dev.github.io/bilby/dynesty-guide.html}.}. We use the same setup throughout this work.

\subsection{GW190814-like injections}
\label{sec:GW190814injections}
To carry out this study, we select a remarkable event in GWTC-2, GW190814, for which strong evidence for higher-order modes was reported~\cite{LIGOScientific:2020zkf, Roy:2019phx}. The source is a compact binary with a highly asymmetric mass ratio, making it particularly suitable for assessing the role of subdominant harmonics. We use the maximum-likelihood sample of \texttt{SEOBNRv4PHM} model released in the GWTC-2 catalog~\cite{LIGOScientific:2020ibl}. We generate injections using \phXPHM model assuming a three-detector network consisting of Hanford (H1), Livingston (L1), and Virgo (V1). To avoid statistical fluctuations from a particular noise realization, the injections are simulated assuming a zero noise realization. In the parameter-estimation runs, the likelihood inner product is weighted by the advanced-design power spectral densities \texttt{aLIGOZeroDetHighPower} for H1 and L1~\cite{aLIGO_ZDHP}, and \texttt{AdvVirgo} for V1~\cite{2012arXiv1202.4031M}, as implemented in \lalsuite. With this setup, the resulting optimal SNR $(\rho_{\rm inj})$ in individual detectors are 28.7, 33.6, and 12.3 in H1, L1, and V1, respectively.
We consider four GW190814-like injections:
\begin{itemize}
\item Case I: the maximum-likelihood sample with $\Gamma = 0$, the resulting network $\rho_{\rm inj} \sim 45.9$;
\item Case II: the maximum-likelihood sample with the primary spin increased to $a_1 = 0.72$, in order to assess the impact of spin effects, and $\Gamma = 0$, the resulting network $\rho_{\rm inj} \sim 48.0$;
\item Case III: the maximum-likelihood sample with \mbox{$\Gamma = 1.5 \times 10^{-3}\,{\rm s^{-1}}$}, the resulting network $\rho_{\rm inj} \sim 45.8$;
\item Case IV: the maximum-likelihood sample with \mbox{$\Gamma = -1.5 \times 10^{-3}\,{\rm s^{-1}}$}, the resulting network \mbox{$\rho_{\rm inj} \sim 45.8$};
\end{itemize}

For the prior setup, we assume a uniform prior on component masses, a uniform prior on individual spin magnitudes with isotropic orientations, and a luminosity distance $d_L$ prior that is uniform in comoving volume. For the acceleration parameter, we assume a uniform prior symmetric around zero.

Fig.~\ref{fig:gw190814_inj12} shows corner plots comparing the recovered marginalized posterior distributions obtained from \losa inference with \phXP versus \phXPHM for an injected \losa (red dashed lines indicate the true parameters). This figure clearly demonstrates that excluding higher harmonics in the recovery waveform leads to systematic biases in the \losa inference. We summarize the results in Table~\ref{tab:GW190814res}.

In all four cases, both the \phXPHM and \phXP analyses show that the recovered posterior of $\Gamma$ is consistent with the injected value. However, the non-inclusion of higher-order modes leads to a considerably broader posterior, $\sim 3 \text{--} 4 \times$ larger. This broadening indicates a loss of sensitivity to small \losa effects when higher harmonics are omitted. Beyond the posterior width, the contour orientations also reveal the underlying parameter degeneracies. In particular, the $(\mathcal{M}_c,\Gamma)$ contours show a strong positive correlation, indicating that part of the \losa--induced phase correction can be partially reabsorbed by a shift in the chirp mass, since both affect the inspiral phasing. The shape of this degeneracy depends on whether higher harmonics are included in the recovery waveform.

For nearly non-spinning injections (Cases I, III, and IV), we also notice significant biases in the recovered luminosity distance, which in turn biases the estimation of source-frame masses. While the high-spin injection (Case II) does not show a noticeable bias in luminosity distance, the posteriors of masses and spins are broader when higher harmonics are omitted. We also note that Cases III and IV, despite having equal-magnitude but opposite-sign injected accelerations, show very similar posterior shapes. This indicates that the dominant behavior is controlled by the degeneracy of $\Gamma$ with the intrinsic phasing parameters rather than by the sign of the injected acceleration alone.

\subsection{GW200105-like eccentric injections}
\label{sec:GW200105injections}
\begin{figure*}[t!]
  \centering%
\includegraphics[width=0.95\linewidth]{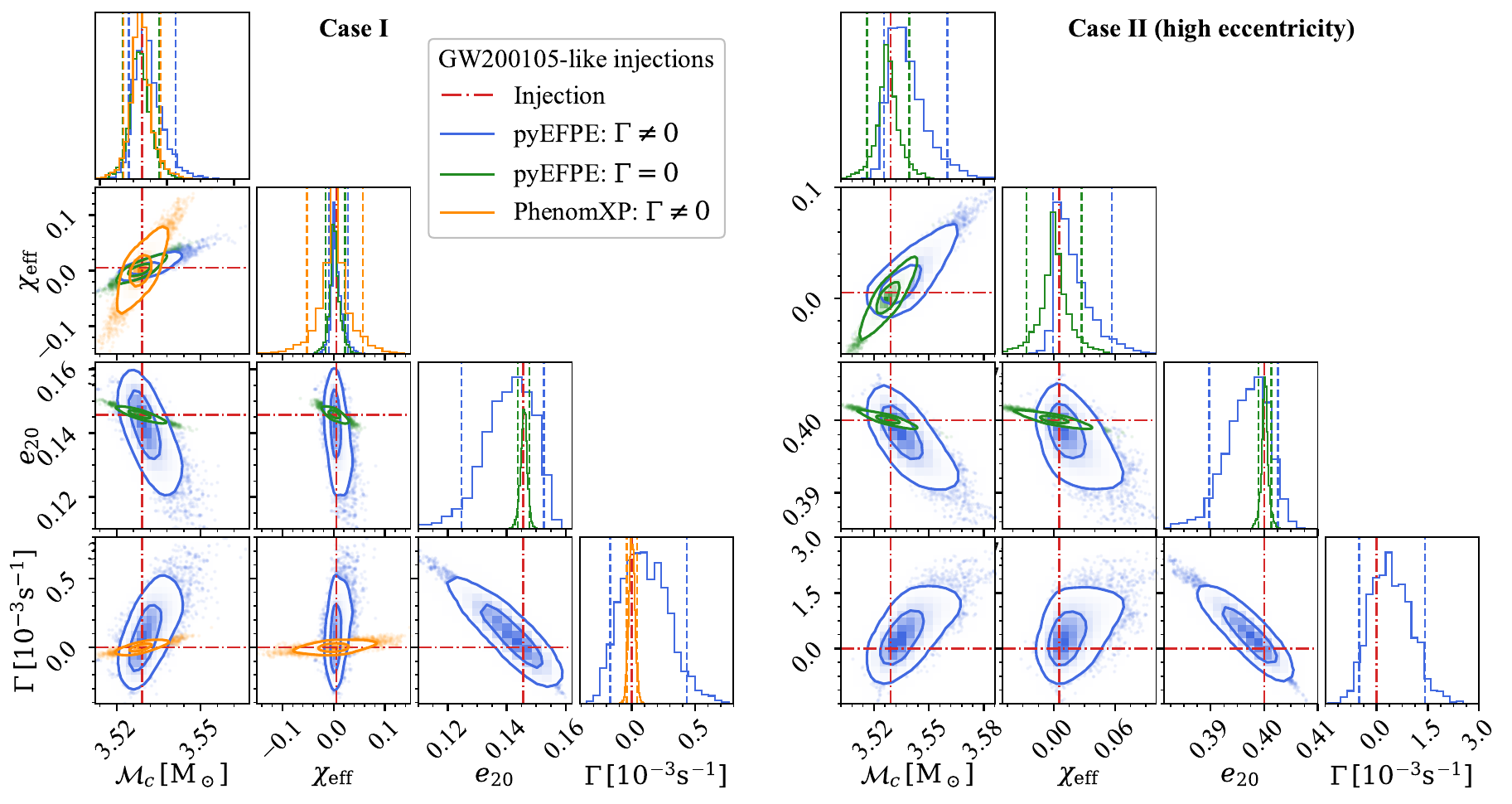}%
  \caption{Illustrating \losa inference for GW200105-like eccentric injections. The corner plot compares the recovered posteriors with \pyEFPE (blue) and \phXP (green) models, highlighting strong $\Gamma$--eccentricity degeneracy. For all these injections, both the injection and recovery models are the same. The red dashed lines indicate the injected values.  The vertical dashed–dotted lines in the one-dimensional histograms mark the 90\% credible interval, and the two contours in the two-dimensional panels enclose the 50\% and 90\% credible regions. The right panel shows the high-eccentricity injection ($e_{20}=0.4$), where the $\Gamma$--eccentricity correlation increases and the \losa measurability degrades, leading to broader posteriors.}
  \label{fig:200105_inj}
\end{figure*}
GW200105 is one of the first confirmed NSBH observations reported by the LVK~\cite{LIGOScientific:2021qlt}. A recent analysis further found strong evidence for orbital eccentricity, inferring a median eccentricity of $\sim 0.145$ at a GW frequency of 20\:\Hz (noted $e_{20}$)~\cite{Morras:2025xfu}, using a eccentric-precessing waveform model \pyEFPE~\cite{Morras:2025xfu}. The presence of eccentricity in this event has also been found in several other analyses using different waveform models~\cite{Planas:2025plq, Kacanja:2025kpr, Jan:2025fps, Clarke:2026cuw, Pompili:2026yxq}. Measurable eccentricity suggests formation history involving dynamical interactions beyond isolated binary evolution, where acceleration is expected to be enhanced by external perturbers. This event is therefore a particularly interesting target for a \losa analysis.

We use the maximum-likelihood sample obtained for this event with the \pyEFPE{} model. As described in the previous section, we perform zero-noise injections for a three-detector network consisting of H1, L1, and V1, assuming their advanced-design sensitivities. To assess the impact of eccentricity on the \losa{} measurement, we consider two types of injection analyses. In the first, both the injection and recovery are performed with the \pyEFPE{} model. In the second, we set the eccentricity of the same maximum-likelihood sample to zero and perform both the injection and recovery with the \phXP{} model. In both cases, we adjust the luminosity distance so that the injected signal has a network SNR of 30.

We keep the injection and recovery models identical in each case in order to isolate the effect of eccentricity on the measurability of \losa{}. Since \pyEFPE{} does not include higher-order modes, we use \phXP{} rather than \phXPHM{} for the quasi-circular comparison.

Fig.~\ref{fig:200105_inj} shows the recovered marginalized posteriors from the \losa inference using the \pyEFPE and \phXP models, with the injected values indicated by the red dashed lines. Although eccentricity enriches the waveform by activating radiation beyond twice the orbital frequency, the recovery of the acceleration parameter is less precise for the eccentric signal than for the quasi-circular one. Thus, in this case, the additional waveform structure does not improve the \losa measurement, but instead leads to a broader posterior on $\Gamma$. The two-dimensional posteriors provide a clearer view of this behavior. In particular, the $(\Gamma,e_{20})$ contours for the eccentric recovery are strongly elongated with a negative slope, showing that larger values of $\Gamma$ can be partially compensated by smaller values of eccentricity. This is the main degeneracy that limits the \losa measurement in the eccentric case.

This behaviour is driven by the strong degeneracy between \losa and eccentricity. The leading-order \losa phase correction enters at $-4$PN, while the leading eccentric correction enters at $-(19/6)$PN, so the two effects appear at nearby PN orders and can partially compensate each other in the waveform This is reflected directly in the orientation of the $(\Gamma,e_{20})$ contours: the likelihood can trade acceleration-induced dephasing against eccentricity-driven dephasing over the finite detector band.  Along the same lines, previous studies of \losa inference using quasi-circular templates for eccentric injections have reported biases and false evidence for acceleration~\cite{Tiwari:2025aec}. Similarly, including the leading-order eccentric phase correction on top of quasi-circular \losa waveforms highlighted the strong degeneracy~\cite{Pathak:2026cik}.

To illustrate the level of this degeneracy, we also perform an additional analysis with $\Gamma=0$, shown in green in Fig.~\ref{fig:200105_inj}. In this case, the posteriors on the eccentricity and chirp mass become significantly narrower than in the analysis with $\Gamma \neq 0$. The comparison with the $\Gamma=0$ analysis further illustrates how the inclusion of \losa broadens the eccentricity and chirp-mass posteriors through correlations with $\Gamma$. Therefore, although eccentric binaries are astrophysically promising targets for \losa{} because they are likely to be hosted in dynamically perturbed environments, but the degeneracy with eccentricity can limit our ability to measure the acceleration reliably.

Looking more closely at the dominant eccentric contribution, at low eccentricity, the frequency-domain inspiral phase can be written as a circular contribution plus an eccentric correction. For the $i$th harmonic, the leading-order phase expressed as~\cite{Yunes:2009yz}
\begin{equation}
\label{eq:ecc_phase_0pn}
\psi_i(f)\simeq
-\frac{3}{128\,\eta}v^{-5}\left(\frac{i}{2}\right)^{8/3}
\left[
1-\frac{2355}{1462}\,e_{\rm ref}^2\, \chi^{-19/3}
\right].
\end{equation}
Here, $e_{\rm ref}\equiv e(f_{\rm ref})$ is the eccentricity defined at a reference frequency $f_{\rm ref}$, and $\chi = v/v_{\rm ref}$, where $v_{\rm ref}\equiv (\pi M f_{\rm ref})^{1/3}$. In the SPA, the stationary time associated with the $i$th harmonic is given as
\begin{equation}
t_i^{\rm SPA}(f) \simeq \frac{5M}{256\,\eta}\,v^{-8}\left(\frac{i}{2}\right)^{8/3}
\left[
1-\frac{157}{43}\,e_{\rm ref}^2 \, \chi^{-19/3}
\right],
\end{equation}
Using the low-eccentricity expansion of the SPA time in Eq.~\eqref{eq:losa_ecc}, the leading-order \losa correction for the $i$th harmonic becomes
\begin{equation}
\Delta\psi_i(f)
\simeq -\frac{25 M\: \Gamma}{65536 \:\eta^2}v^{-13} \left(\frac{i}{2}\right)^{16/3}\left[
1-\frac{157}{43}\,e_{\rm ref}^2\, \chi^{-19/3}
\right]^2 \, .
\label{eq:losa_ecc_low}
\end{equation}
Both the eccentric correction to the inspiral phasing in Eq.~\eqref{eq:ecc_phase_0pn} and the \losa correction are dominated by the lowest-frequency portion of the signal: the eccentric contribution grows toward low frequency through the velocity order $v^{-34/3}$, while the \losa phase shift enters at $v^{-39/3}$. As a result, over a finite detector band, the likelihood can partially trade off a nonzero $\Gamma$ against eccentricity-driven dephasing, producing strong posterior correlations between $\Gamma$ and $e_{\rm ref}$ when eccentric features are not measured with sufficient precision.

At moderate eccentricity, the eccentricity-dependent factor in Eq.~\eqref{eq:losa_ecc_low} becomes more prominent, amplifying the correlation between $\Gamma$ and $e_{\rm ref}$ in the inference. At the same time, eccentricity accelerates the intrinsic inspiral, reducing the time-to-coalescence at a given frequency and thereby decreasing the overall magnitude of the \losa phase shift, since $\Delta\psi_i \propto f\,(t_i^{\rm SPA})^2$. Consequently, increasing eccentricity can both amplify the $\Gamma\text{--}e_{\rm ref}$ degeneracy and reduce the net \losa contribution to phase, leading to a broader posterior on $\Gamma$ and degraded \losa measurability in high-eccentricity injections, as shown in the right panel of Fig.~\ref{fig:200105_inj}.

\begin{figure}[t]
  \centering%
\includegraphics[width=\linewidth]{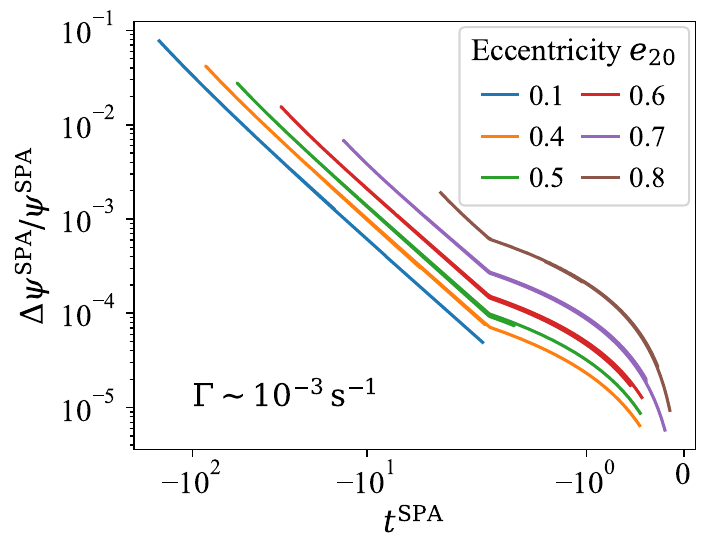}%
  \caption{Fractional \losa{}-induced phase shift $\Delta\psi^{\rm SPA}/\psi^{\rm SPA}$, computed as a function of the stationary time $t^{\rm SPA}$ for a set eccentricities $e_{20}$ and fixed acceleration $\Gamma \sim 10^{-3}$. We assumed an equal mass non-spinning binary with component masses $1.4\:\msun$.}
  \label{fig:losa_ecc_relphase}
\end{figure}

Eq.~\eqref{eq:losa_ecc_low} suggests a non-monotonic dependence on eccentricity: increasing $e_{\rm ref}$ initially reduces the \losa correction through the negative $e_{\rm ref}^2\chi^{-19/3}$ contribution, while for sufficiently large $e_{\rm ref}$, it could eventually increase again. However, for moderately to highly eccentric signals, the dominant effect is the change in the SPA time itself. As eccentricity increases, the inspiral proceeds more rapidly because of the enhanced dissipation near periastron, so the stationary time-to-coalescence $t^{\rm SPA}$ at a given frequency decreases in magnitude. Since the \losa phase correction is proportional to $t^{2}(f)$, this reduction suppresses the net \losa effect even when the eccentricity-dependent factor in Eq.~\eqref{eq:losa_ecc_low} becomes sizable. This behavior is illustrated in Fig.~\ref{fig:losa_ecc_relphase}, where we compute the fractional phase shift $\Delta\psi^{\rm SPA}/\psi^{\rm SPA}$ directly from \pyEFPE{} for a sequence of eccentricities. The fractional dephasing decreases with increasing eccentricity, implying that the reduction of the SPA time-to-coalescence in highly eccentric inspirals suppresses the net \losa effect.


\section{Analysis of real events}
\label{sec:real_events}
We analyze a subset of LVK events selected for their potential relevance to \losa effects. In particular, we consider three events observed during O3: GW190814~\cite{LIGOScientific:2020zkf}, GW200105~\cite{LIGOScientific:2021qlt}, and GW190728\_064510 (hereafter GW190728)~\cite{LIGOScientific:2020ibl}. For this analysis, we use the data released in the GWTC-2.1 and GWTC-3 catalogs by LVK~\cite{LIGOScientific:2021usb, KAGRA:2021vkt, KAGRA:2023pio}. We follow the parameter-estimation configuration used in the corresponding LVK analyses.

\begin{figure}[t]
  \centering%
\includegraphics[width=\linewidth]{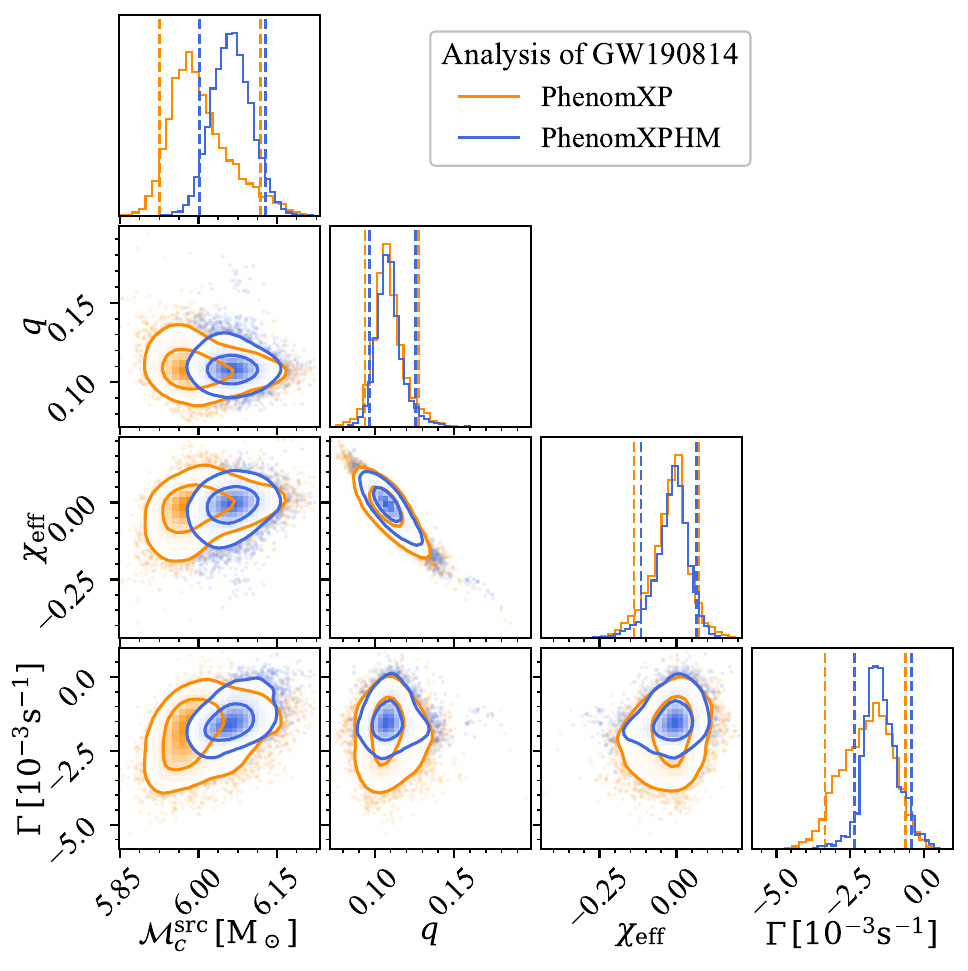}%
  \caption{Marginalized posterior distributions from the \losa analyses of GW190814 signal, using \phXP (orange) and \phXPHM (blue) models. The vertical dashed lines in 1D
histograms represent the 90\% credible interval. The two consecutive contours in the 2D joint plot correspond to 50\% and 90\% credible regions.}
  \label{fig:gw190814_losa}
\end{figure}

\begin{figure}[t]
  \centering%
\includegraphics[width=\linewidth]{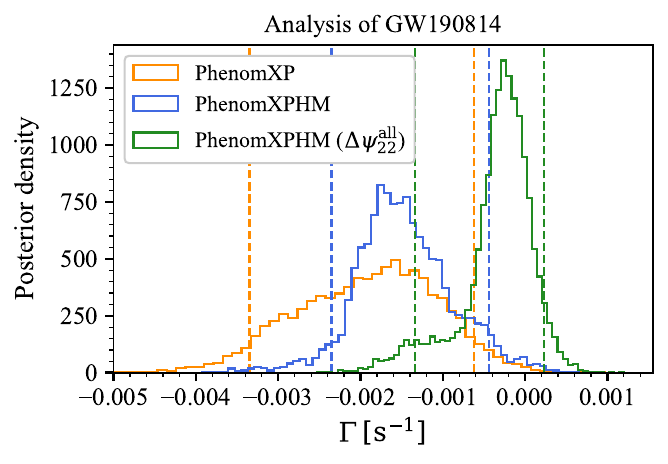}%
  \caption{Comparison of the marginalized posterior distributions for the acceleration parameter $\Gamma$ from different analyses of GW190814 using various waveform models. The green histogram labeled \phXPHM ($\Delta \psi_{22}^{\rm all}$) shows the \losa result obtained with the \phXPHM model when the \losa correction is computed using a quadrupolar prescription and applied uniformly across the full higher-harmonic waveform, similar to Ref.~\cite{Hendriks:2026kys}. The inconsistent application of \losa correction in the higher-order modes can lead to a reduction in the support for a non-zero $\Gamma$.}
  \label{fig:gw190814_gamma}
\end{figure}

\subsection{Analysis of GW190814}
\label{sec:GW190814analyses}
A recent study by Yang et al.~\cite{Yang:2024tje} reported evidence for \losa in GW190814, with an inferred 90\% credible interval of \mbox{ $\Gamma \sim 1.5^{+0.8}_{-0.8}\times 10^{-3}\:\invsec$}, and interpreted this as suggesting the presence of a nearby compact object. A subsequent study by Hendriks et al. repeated the analysis and found no support for \losa in GW190814~\cite{Hendriks:2026kys}, showing that the apparent evidence disappears when the signal is analyzed using a sufficiently long data segment. This discrepancy motivates a reanalysis of the signal using updated waveform models. For the prior setup, we broaden the chirp-mass range to $\mathcal{M}_c\in [5,7]\:\msun$ to avoid railing effects that may arise from the degeneracy with acceleration.

Fig.~\ref{fig:gw190814_losa} shows the results for GW190814, comparing the posteriors inferred using the \phXP and \phXPHM models. The recovered values for acceleration are \mbox{$-1.84^{+1.26}_{-1.46} \times 10^{-3} \:\invsec$} and \mbox{$-1.51^{+1.12}_{-0.78} \times 10^{-3} \:\invsec$}, respectively. For both models, $\Gamma=0$ lies outside the 90\% credible interval, but remains within the posterior support. To quantify the significance of this offset, we compute the fraction of the posterior enclosed by the smallest highest-posterior-density (HPD) interval that contains the vacuum value, $\Gamma=0$. We find that the vacuum value is contained within the 98.7\% and 95.6\% HPD intervals, respectively. We also note that the logarithmic Bayes factors for the \losa hypothesis relative to the vacuum hypothesis are close to zero: specifically, $\Delta\log_{10}\mathcal{B}=0.52$ and $0.03$, respectively. We therefore conclude that the evidence supporting a \losa effect is not significant.

To illustrate the role of higher-order modes in the \losa inference, we also compare the recovered posteriors for the source-frame chirp mass $\mathcal{M}_c^{\rm src}$, $q$, and $\chi_{\rm eff}$ between the \phXP and \phXPHM analyses. Among these parameters, the most noticeable change appears in $\mathcal{M}_c^{\rm src}$, while the posteriors for $q$ and $\chi_{\rm eff}$ remain broadly consistent between the two models. This behavior is consistent with a biased inference of $d_L$ when higher-order modes are omitted in the recovery waveform, which propagates into the source-frame mass conversion. Overall, this trend is consistent with the \mbox{Case I} injection study presented in Sec.~\ref{sec:GW190814injections}.

We now comment on the apparent tension between the \losa claim by Yang et al.~\cite{Yang:2024tje} and the null result reported by Hendriks et al.~\cite{Hendriks:2026kys}. Yang et al. report a nonzero \losa posterior for GW190814 and a Bayes factor value of 58/1 in favor of the \losa hypothesis over the isolated case. Hendriks et al. show that the reported \losa evidence disappears when the analysis is repeated with a sufficiently long data segment, and they identify the short signal duration used in the original setup as the main driver of the discrepancy.
A further modeling aspect concerns how the \losa correction is implemented when using a higher-mode, precessing waveform model \phXPHM. In both Yang et al.~\cite{Yang:2024tje} and Hendriks et al.~\cite{Hendriks:2026kys}, the \losa modification is derived using a quadrupolar phasing prescription and then applied to the full multipolar waveform, without an explicit mode-by-mode \losa treatment for the higher harmonics. When the contribution of higher-order modes is non-trivial, this approximation can introduce systematic biases, as shown in the previous section. To test this explicitly, we repeat the \phXPHM analysis, where \losa correction is calculated using a quadrupolar prescription and applied uniformly across the full higher-harmonic waveform. We use an analysis duration of 32 s, following the LVK configuration~\cite{LIGOScientific:2020zkf}. 

Fig.~\ref{fig:gw190814_gamma} compares the resulting $\Gamma$ posteriors from three different analyses. It shows that this approximation shifts the inference toward $\Gamma=0$ and exhibits a narrower 90\% credible interval, indicating significant biases. In addition, we find that substantial biases can arise if the \losa{} correction is applied only to the $(2,2)$ mode while higher-order modes remain unmodified. As discussed further in Appendix~\ref{app:GW190814Injections}, such inconsistent treatments can bias not only the recovered acceleration but also other source parameters through spurious correlations, ultimately reducing the support for a non-zero \(\Gamma\). Our results therefore highlight that a consistent mode-by-mode \losa{} treatment is crucial for robust inference when higher harmonics are relevant.

\subsection{Analysis of GW200105}
This event was observed by two detectors, L1 and V1, while H1 was not operational at that time, with a network SNR of $\sim 13.6$. It is effectively dominated by a single detector (L1)~\cite{LIGOScientific:2021qlt}. In addition, light-scattering noise was identified in L1 data below $25\Hz$, and the strain data was cleaned through a glitch-subtraction procedure. The Virgo data affected by systematic calibration errors in the frequency window $46\text{--}51\Hz$ was excluded in the LVK analysis by masking the power spectral density in that range. Following~\cite{Morras:2025xfu}, we compute the likelihood in a frequency range $20\text{--}280\Hz$. We perform the \losa analysis using the modified \pyEFPE model with a uniform prior on eccentricity, $e_{20} \in [0, 0.4]$.
\begin{figure}[t]
  \centering%
\includegraphics[width=\linewidth]{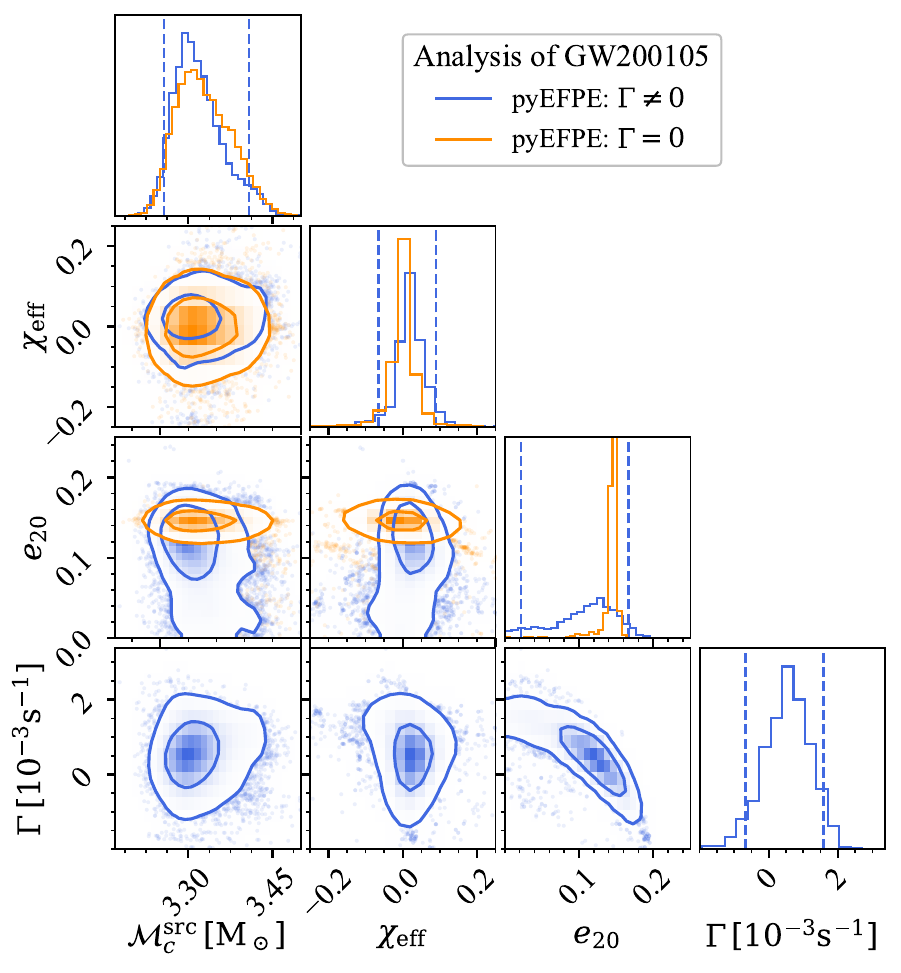}%
  \caption{Marginalized posterior distributions from the \losa analyses of GW200105 signal using \pyEFPE model. The vertical dashed lines in 1D histograms represent the 90\% credible interval. The two consecutive contours in the 2D joint plot correspond to 50\% and 90\% credible regions. Highlights strong degeneracy between eccentricity and acceleration.}
  \label{fig:gw200105_losa}
\end{figure}

Fig.~\ref{fig:gw200105_losa} shows the results of the \losa analysis for GW200105. The recovered acceleration parameter is
\mbox{$\Gamma \sim 0.56^{+1.01}_{-1.23} \times 10^{-3}\:\invsec$} at 90\% credibility, indicating no evidence for \losa. We also show the marginalized distributions for $\mathcal{M}_c$, $\chi_{\rm eff}$, and $e_{20}$. The 2D plot of $\Gamma$--$e_{20}$ shows strong degeneracy between them. The higher value of acceleration can compensate the effect of orbital eccentricity such that support for eccentricity decreases considerably. This behaviour is consistent with the zero-noise injection studies as demonstrated in Sec.~\ref{sec:GW200105injections}.

Eccentricity enriches the waveform by activating radiation beyond the dominant quadrupolar harmonic. While this additional structure can help constrain some beyond-GR effects~\cite{Roy:2025xih}, in the present case eccentricity correlates strongly with \losa{} and degrades the measurability of $\Gamma$.

\subsection{Analysis of GW190728}
A recent study reported tentative evidence for a scalar-field environment around the source of GW190728~\cite{Roy:2025qaa}. The leading-order phase correction from such an environment enters at a large negative PN order. Since several other astrophysical environmental effects also contribute at large negative PN order, the reported signature could, in principle, be partially degenerate with \losa. That study further found no substantial evidence for \losa when analyzing GW190728 with the \phXP model.
\begin{figure}[t]
  \centering%
\includegraphics[width=\linewidth]{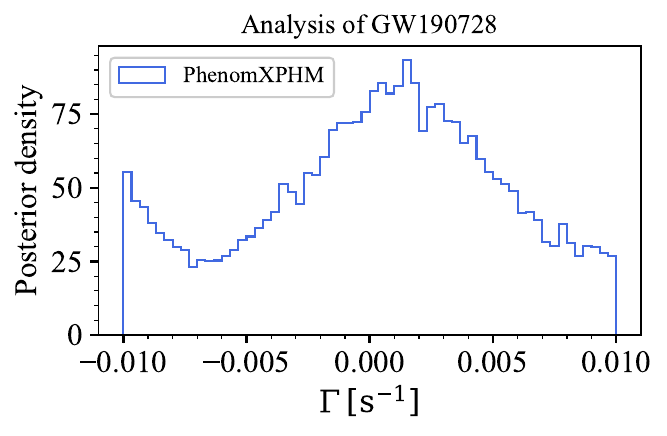}%
  \caption{Marginalized posterior distributions for the acceleration parameter $\Gamma$ from the analyses of GW190728 using \phXPHM model, indicating uninformative posterior and \losa is not constrained.}
  \label{fig:gw190728_gamma}
\end{figure}

Motivated by our injection results showing that an inconsistent treatment of higher harmonics can bias \losa inference, we reanalyze GW190728 using the \phXPHM model. We adopt the same prior on the acceleration parameter, \mbox{$\Gamma \in [-0.01, +0.01]\,\invsec$}, used throughout this work. We find that the data do not yield an informative constraint on $\Gamma$, as shown in Fig.~\ref{fig:gw190728_gamma}.
The increase in posterior density near the lower prior boundary indicates a prior-railing effect. To investigate this further, we repeat the analysis using a broader prior on $\Gamma$. The resulting posterior peaks at a larger-magnitude negative value of $\Gamma$, while retaining nonzero posterior support at $\Gamma=0$. Although large accelerations may arise for compact binaries close to SMBHs or in AGN disks~\cite{Peng:2021vzr, Zwick:2025wkt, Tagawa:2025tfd}, the preferred values of $\abs{\Gamma}$ are well above the characteristic ranges predicted for most of the astrophysical populations considered in these studies. We therefore do not expect \losa to account for the previously reported scalar-field environmental signature in this event.

\section{Conclusion}
\label{sec:conclusion}
In this work, we revisited line-of-sight acceleration (\losa) as an environmental signature in compact-binary GW signals. We extended the standard \losa treatment beyond the dominant quasi-circular quadrupole by implementing the phase (and leading-order amplitude) corrections in a mode-by-mode manner for higher-harmonic waveform models, and by incorporating it in a precessing and higher-order mode model (\phXPHM) and in an eccentric inspiral model (\pyEFPE). Our formulation is based on the integrated time-delay interpretation, which allows the \losa correction to be expressed directly in terms of the stationary-phase time and applied consistently to each contributing harmonic.

For quasi-circular injections, we find that omitting higher-order modes in the recovery waveform does not systematically shift the inferred acceleration away from the injected value. Instead, the main effect is a loss of constraining power: the posterior on $\Gamma$ becomes significantly broader, reducing sensitivity to small \losa effects. We also observe that neglecting higher-order modes can bias the luminosity-distance inference, which propagates into derived source-frame quantities and leads to biased estimates of the source-frame chirp mass. For the highly spinning binary, the luminosity-distance bias is reduced when higher-order modes are neglected, but the posteriors on the masses and spins remain significantly broader. Overall, an inconsistent treatment of higher-order modes can degrade \losa measurability and bias source-frame mass estimates, even when it does not produce a spurious nonzero acceleration.

For eccentric injections, we find that allowing both eccentricity and \losa in the inference increases degeneracies and broadens the posterior on the acceleration parameter. In particular, for high eccentricity, we show that the fractional \losa{}-induced phase shift becomes smaller, consistent with the reduced stationary time-to-coalescence in highly eccentric inspirals. This explains why \losa measurability can worsen as eccentricity increases, despite the richer waveform morphology.

Finally, we analyzed a set of selected LVK events from O3 that are potentially relevant for \losa studies, including GW190814, GW200105, and GW190728. Across these events, we find no statistically significant evidence supporting a nonzero \los acceleration. In the case of GW190814, our results further highlight that consistent mode-by-mode treatment of \losa in the presence of higher harmonics is crucial for robust inference and for resolving discrepancies between different analyses in the literature.

This new framework provides a practical route to extend \losa searches to larger GW catalogs and to waveform models with richer signal morphology, including higher harmonics and eccentricity. As detector sensitivity improves and inspirals can be observed for longer, the cumulative Doppler-induced phase distortions will be measured with higher precision, making \losa an increasingly useful observable to probe dynamical environments and hierarchical perturbations. More broadly, consistent mode-by-mode implementations will be essential for avoiding spurious environmental inferences and for enabling robust population-level studies that connect GW measurements to compact-binary formation channels.

\section{Code availability}
The repository containing the waveform model and the scripts used to reproduce the figures in this paper will be made available at Ref.~\cite{Roy2026gwLoSA}.

\section{acknowledgments}
Computational resources have been provided by the supercomputing facilities of the Universit{\'e} catholique de Louvain (CISM/UCL) and the Consortium des Équipements de Calcul Intensif en F{\'e}d{\'e}ration Wallonie Bruxelles (C{\'E}CI) funded by the Fond de la Recherche Scientifique de Belgique (F.R.S.-FNRS) under convention 2.5020.11 and by the Walloon Region. We have used \numpy~\cite{Harris:2020xlr}, \scipy~\cite{Virtanen:2019joe}, \matplotlib~\cite{Hunter:2007ouj} for analyses and preparing the figures in the manuscript. This research has made use of data or software obtained from the Gravitational Wave Open Science Center (gwosc.org), a service of the LIGO Scientific Collaboration, the Virgo Collaboration, and KAGRA. This material is based upon work supported by NSF's LIGO Laboratory which is a major facility fully funded by the National Science Foundation, as well as the Science and Technology Facilities Council (STFC) of the United Kingdom, the Max-Planck-Society (MPS), and the State of Niedersachsen/Germany for support of the construction of Advanced LIGO and construction and operation of the GEO600 detector. Additional support for Advanced LIGO was provided by the Australian Research Council. Virgo is funded, through the European Gravitational Observatory (EGO), by the French Centre National de Recherche Scientifique (CNRS), the Italian Istituto Nazionale di Fisica Nucleare (INFN) and the Dutch Nikhef, with contributions by institutions from Belgium, Germany, Greece, Hungary, Ireland, Japan, Monaco, Poland, Portugal, Spain. KAGRA is supported by Ministry of Education, Culture, Sports, Science and Technology (MEXT), Japan Society for the Promotion of Science (JSPS) in Japan; National Research Foundation (NRF) and Ministry of Science and ICT (MSIT) in Korea; Academia Sinica (AS) and National Science and Technology Council (NSTC) in Taiwan. 
\onecolumngrid
\clearpage
\begin{figure*}[h]
\centering
  \includegraphics[width=0.95\textwidth]{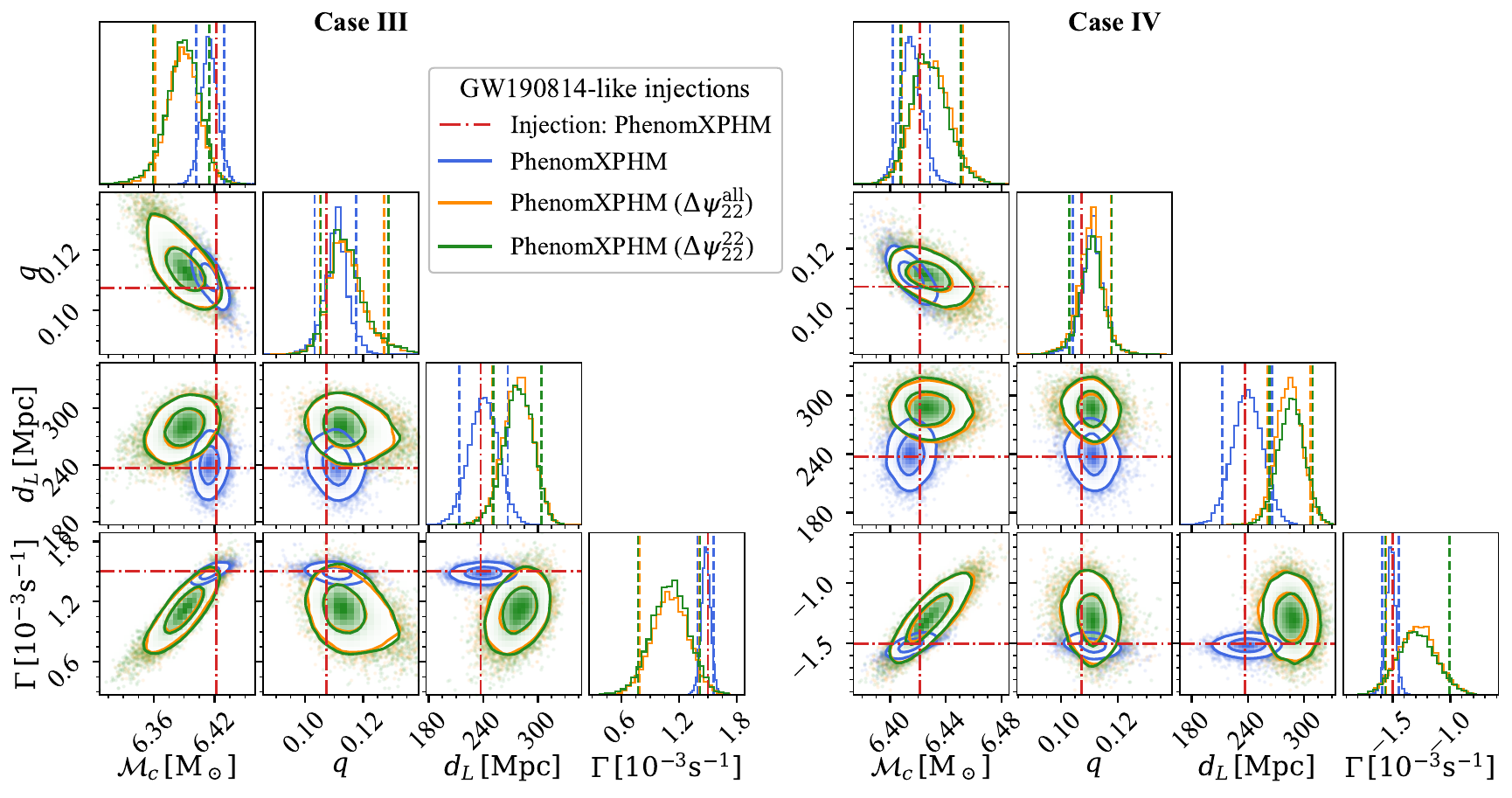}
  \caption{Systematic biases in \losa inference due to inconsistent treatment of higher harmonics in the recovery waveform, shown for GW190814-like injections. The injections are generated using the \phXPHM model as described in Sec.~\ref{sec:GW190814injections}. The red dashed lines indicate the injected values. The vertical dashed-dot lines in 1D histograms represent the 90\% credible interval. The two consecutive contours in the 2D joint plot correspond to 50\% and 90\% credible regions. For details, see companion text in Appendix.~\ref{app:GW190814Injections}.}
  \label{fig:GW190814_injs_appendix}
\end{figure*}

\twocolumngrid
\appendix

\section{Inconsistent treatment of higher harmonics (GW190814-like injections) }
\label{app:GW190814Injections}

We further highlight the impact of an inconsistent treatment of higher-order modes when incorporating \losa corrections into the baseline waveform model. We adopt the same injection setup as described in Sec.~\ref{sec:GW190814injections} and focus on Cases III and IV, for which $\Gamma \neq 0$. As discussed in Sec.~\ref{sec:GW190814analyses}, we consider two approximate prescriptions. In the first, the \losa correction is computed using the quadrupolar (2,2) mode and applied uniformly to all modes of \phXPHM, denoted by \phXPHM ($\Delta\psi_{22}^{\rm all}$). In the second, the \losa correction constructed from the $(2,2)$ mode is applied only to the (2,2) mode itself, denoted by \phXPHM ($\Delta\psi_{22}^{22}$).

Figure~\ref{fig:GW190814_injs_appendix} shows the recovered posteriors obtained with these two approximate prescriptions, compared to the fully consistent \phXPHM implementation as also shown in Fig.~\ref{fig:gw190814_inj12}. In both cases, the inconsistent treatment of higher-order modes produces substantial biases in the recovered source parameters. The effect is particularly pronounced for the acceleration parameter $\Gamma$, whose posterior is shifted toward smaller $\abs{\Gamma}$ and significantly away from the injected value. Consequently, the support for a non-zero acceleration is substantially reduced. The inconsistent prescriptions also introduce strong correlations between $\Gamma$ and the intrinsic binary parameters. In particular, the $(\mathcal{M}_c,\Gamma)$ contours become strongly elongated, indicating that part of the incorrectly modelled \losa phasing is absorbed through shifts in the chirp mass. For positive (negative) injected accelerations, this leads to a systematic shift of the recovered chirp mass toward smaller (larger) values.

Significant biases are also observed in the luminosity distance and mass-ratio posteriors. In both Cases III and IV, the recovered $d_L$ posteriors are displaced from the injected value, and the corresponding two-dimensional contours are noticeably offset from the true source parameters. The two approximate prescriptions yield remarkably similar results, indicating that the dominant source of bias is the use of a quadrupolar \losa correction in a waveform containing higher-order modes. These results demonstrate that incorporating higher-order modes alone is not sufficient for robust \losa inference; the \losa correction must also be applied consistently across the harmonic structure of the waveform. Otherwise, significant parameter biases can arise and the evidence for acceleration can be substantially weakened.


\bibliographystyle{apsrev4-2-titles}
\bibliography{reference}

\end{document}